\documentclass{aa}   
\usepackage{graphicx}
\usepackage[usenames, dvipsnames]{color}
\usepackage{txfonts}
\usepackage[colorlinks=true,linkcolor=blue,citecolor=blue,urlcolor=blue]{hyperref}
\usepackage{soul}
\usepackage[normalem]{ulem}
\newcommand{\h}{H$_{2}$}
\newcommand{\kms}{km s$^{-1}$}
 
\begin{document} 


\title{Clumpiness of the interstellar medium in the central parsec of the Galaxy from H$_{2}$ flux--extinction correlation}
\author{
        A. Ciurlo\inst{1}\inst{2}, T. Paumard\inst{1}, D. Rouan\inst{1} \and Y. Cl\'enet\inst{1}
          }
\institute{
LESIA, Observatoire de Paris, PSL Research University, CNRS, Sorbonne Universit\'es, UPMC Univ. Paris 06, Univ. Paris Diderot, Sorbonne Paris Cit\'e, 5 place Jules Janssen, 92195 Meudon, France\\  \email{ciurlo@astro.ucla.edu; thibaut.paumard@obspm.fr; daniel.rouan@obspm.fr 
} \and University of California Los Angeles, 430 Portola Plaza, Los Angeles, CA 90095, USA \\ 
             }

\date{}

\titlerunning{Molecular gas clumpiness in the Galactic Center}


\abstract
{The central parsec of the Galaxy contains a young star cluster embedded in a complex interstellar medium. 
The latter mainly consists of a torus of dense clumps and streams of molecular gas (the circumnuclear disk, CND) enclosing streamers of ionized gas (the Minispiral).}
{In this complex environment, knowledge of the local extinction that locally affects each feature is crucial to properly study and disentangle them.
We previously studied molecular gas in this region and inferred an extinction map from two H$_{2}$ lines. 
Extinction appears to be correlated with the dereddened flux in several contiguous areas in the field of view.
Here, we discuss the origin of this local correlation.} 
{We model the observed effect with a simple radiative transfer model. 
\h~emission arises from the surfaces of clumps (i.e., shells) that are exposed to the ambient ultraviolet (UV) radiation field. 
We consider the shell at the surface of an emitting clump. The shell has a varying optical depth and a screen of dust in front of it. 
The optical depth varies from one line of sight to another, either because of varying extinction coefficient from the shell of one clump to that of another or because of a varying number of identical clumps on the line of sight.}
{In both scenarios, the model accurately reproduces the dependence of molecular gas emission and extinction.
The reason for this correlation is that, in the central parsec, the molecular gas is mixed everywhere with dust that locally affects the observed gas emission.
In addition, there is extinction due to foreground (`screen') dust.}
{This analysis favors a scenario where the central parsec is filled with clumps of dust and molecular gas.
Separating foreground from local extinction allows for a probe for local conditions (H$_{2}$ is mixed with dust) and can also constrain the three-dimensional (3D) position of objects under study. }


\maketitle
   
\section{Introduction}
\label{intro}

The central parsec of the Galaxy hosts a supermassive black hole (Sgr~A*, \citealt{Lynden-Bell71}). 
In the immediate environment of Sgr~A* there is a very young star cluster \citep{Becklin68} emitting strong ultraviolet (UV) radiation \citep{Martins07}. 
This region exhibits two main structures of interstellar medium (ISM): the circumnuclear disk (CND), composed of dust and neutral gas \citep{Becklin82}, and the \ion H{ii} region Sgr~A~West (the Minispiral), consisting of dust and ionized gas \citep{Lo83}. 
The ISM, and particularly its molecular phase, plays an important role in Galaxy evolution, fueling star formation and the growth of super massive black holes. 

At optical wavelengths, the Galactic center is obscured by an extinction of A$_{V}\sim$30 magnitudes because of diffuse dust along the line of sight \citep{Rieke89}. 
Later works \citep{Fritz11,Gao13} suggest that the extinction might be even higher. 
To properly analyze any component (star, gas, dust) in terms of photometry or morphology, extinction correction is crucial.
In the past decade, various authors have addressed the question of the local variations of the extinction in the Galactic center and have traced the extinction map of this region using different tracers \citep[among others]{Scoville03,Schodel10,Fritz11}. 

Two important aspects must be considered: the behavior of the extinction law as a function of the wavelength and the local variations of the spectral extinction coefficient. 
These local variations of the extinction are introduced by the specific environment to which the emitter is subjected. 
Local extinction must be estimated to distinguish between variations of brightness caused by extinction and by the intrinsic source.       
                
Even though the bulk of extinction is foreground and amounts to 2.5~mag at K band \citep{Fritz11}, most of the variation in extinction is local. 
For instance \cite{Paumard04} showed that the dust contained in the Eastern Bridge of the Minispiral is responsible for 0.76~mag of extinction at K of the Northern Arm. 
\cite{Schodel10} obtained a A$_{Ks}$ map, showing local variations reaching 1 mag, with a typical dispersion of 0.5 mag.

One must point out that extinction correction of an object is dependent on its three-dimensional (3D) position. 
Extinction maps are of limited use because of the fact that the various emitters -- stars, ionized gas, and molecular gas -- are not at the same 3D position and not at the same line of sight position with respect to the absorbing dust.
This is also true for any source used as a probe to evaluate the local extinction.
For this reason it is better, when possible, to evaluate the extinction variations directly from the emitter that is being studied. 
        
In \cite{Ciurlo16} we analyzed H$_{2}$ in the central parsec. 
To do so, we used a spectro-imagery datacube obtained by SPIFFI \citep{Tecza00} at the VLT in 2003 \citep{Eisenhauer03}.
The observed field of view is $36\arcsec\times29\arcsec$, with a spectral resolution $R = 1\,300$ in the near-infrared (NIR). 
The spectral range of the observation covers several H$_{2}$ lines.

To properly describe the gas morphology we built an extinction map from two of the observed lines.  
We considered 1-0 S(1) and 1-0 Q(3) transitions at 2.1218~$\mu$m and 2.4236~$\mu$m, respectively, which correspond to transitions arising from the same upper level.
We derived the extinction map and used it to deredden the flux map of the H$_{2}$ line 1-0 S(1).
The description of the dataset, the method used, the derived extinction and the whole procedure can be found in \cite{Ciurlo16}.
In the current paper, we concentrate on the relation between the extinction and line flux maps to evaluate the effects of local extinction.
        
In Section~\ref{obs} we show that the dereddened flux map and the extinction map show local correlations. 
In Section~\ref{model} we propose a phenomenological model to account for the observed effects.
This model describes the observations well, as we show in Section~\ref{comparison}.
In Section~\ref{conclusions} we discuss how this analysis shows that H$_{2}$ is everywhere mixed with dust in the central cavity of the CND, where it is arranged in multiple (up to 20) small-scale clumps.
        
\section{Observation of the flux--extinction correlation} 
\label{obs}     

                \begin{figure}
                \centering
                \includegraphics[width=9cm]{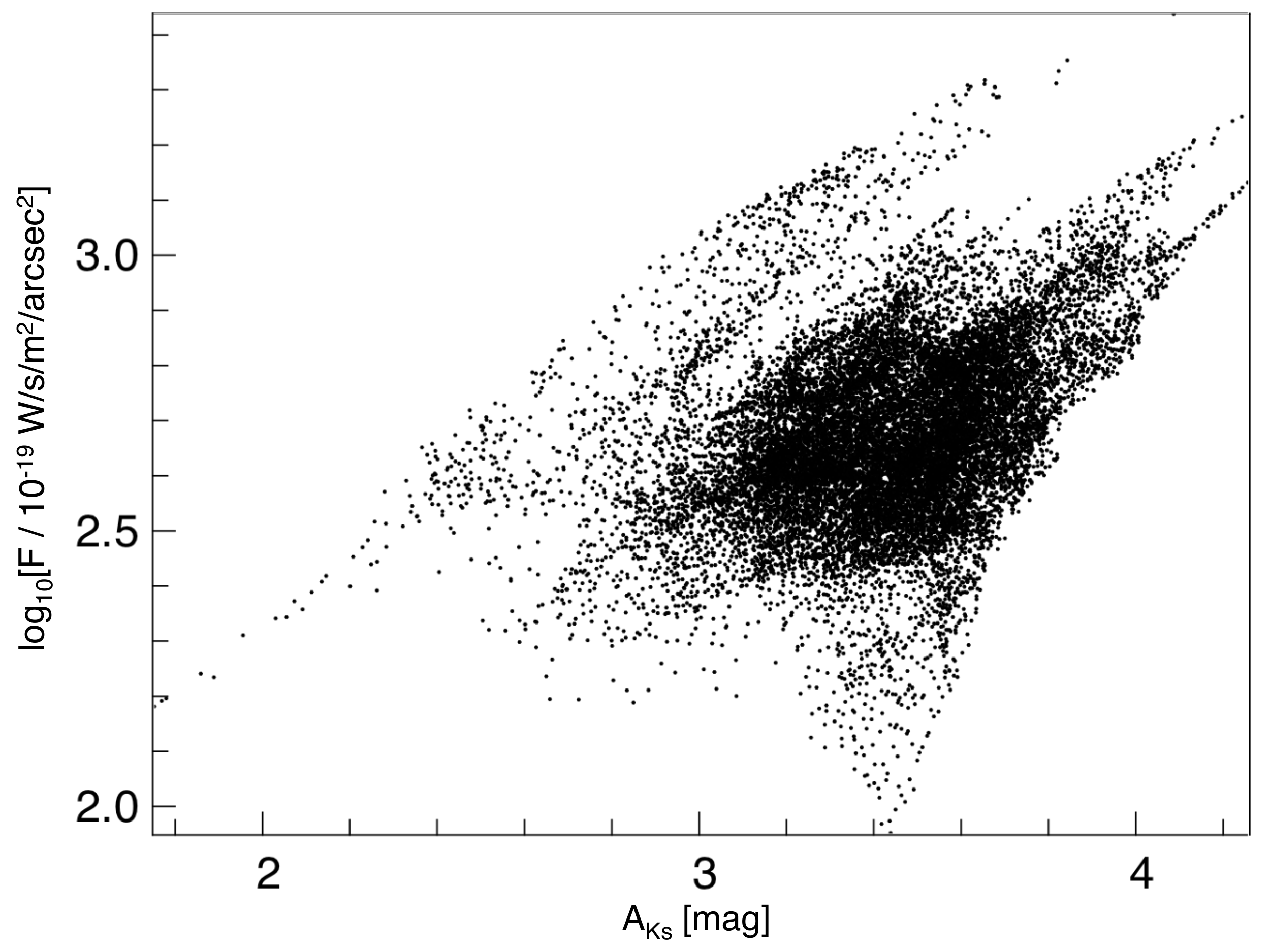}
                \caption{H$_{2}$ 1-0~S(1) line dereddened flux against A$_{Ks}$ for each pixel in the SPIFFI field of view.}
                \label{res:corr}
                \end{figure}

In \cite{Ciurlo16} the extinction A$_{Ks}$ is derived by simply assuming that the primary source is an optically thin layer of hot \h. 
This is assuming that the ratio of the primary fluxes of lines Q(3) and S(1) is simply given by their Einstein coefficients and wavelengths ($\mathcal{A}_{Q3} / \mathcal{A}_{S1} \cdot \lambda_{S1} /  \lambda_{Q3}$).
The obtained dereddened flux map and A$_{Ks}$ map (\citealt{Ciurlo16}, Fig.~5) appear correlated: the flux is high where extinction is high.           
In order to check this apparent correlation we plotted the logarithm of the dereddened flux of the 1-0~S(1) line against A$_{Ks}$ for each pixel in the field (Fig.~\ref{res:corr}). 
A correlation between flux and A$_{Ks}$ would appear as a continuous monotonically rising line on this plot. 
However, the plot does not show a single line, but rather several elongated groups of data points.  
One can distinguish at least four sets of groups of points in the correlation plot, as if there were no global correlation, but instead several local correlations.
        
\begin{figure}
                \centering
                \includegraphics[width=9.cm]{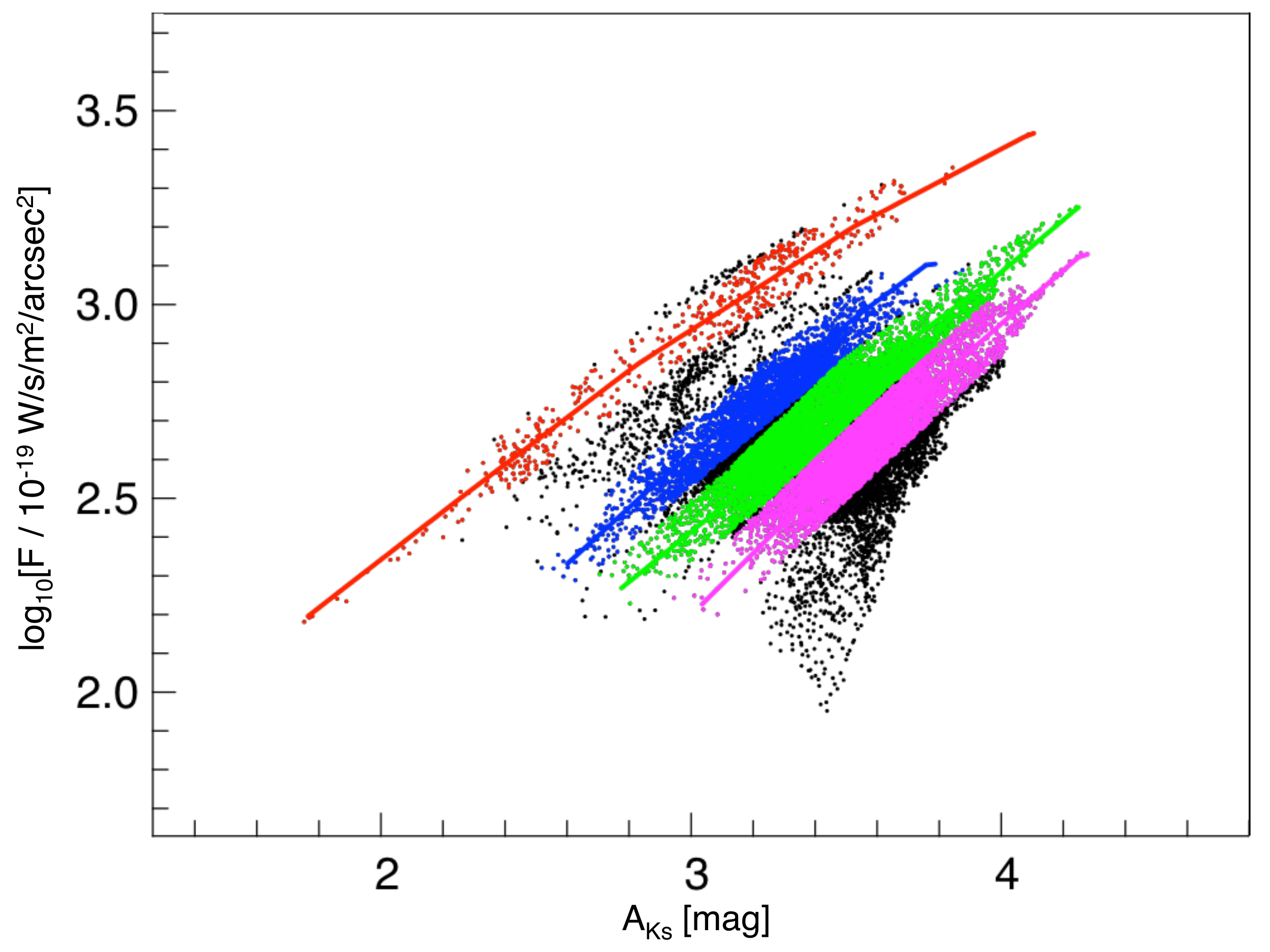}
                \includegraphics[width=8cm]{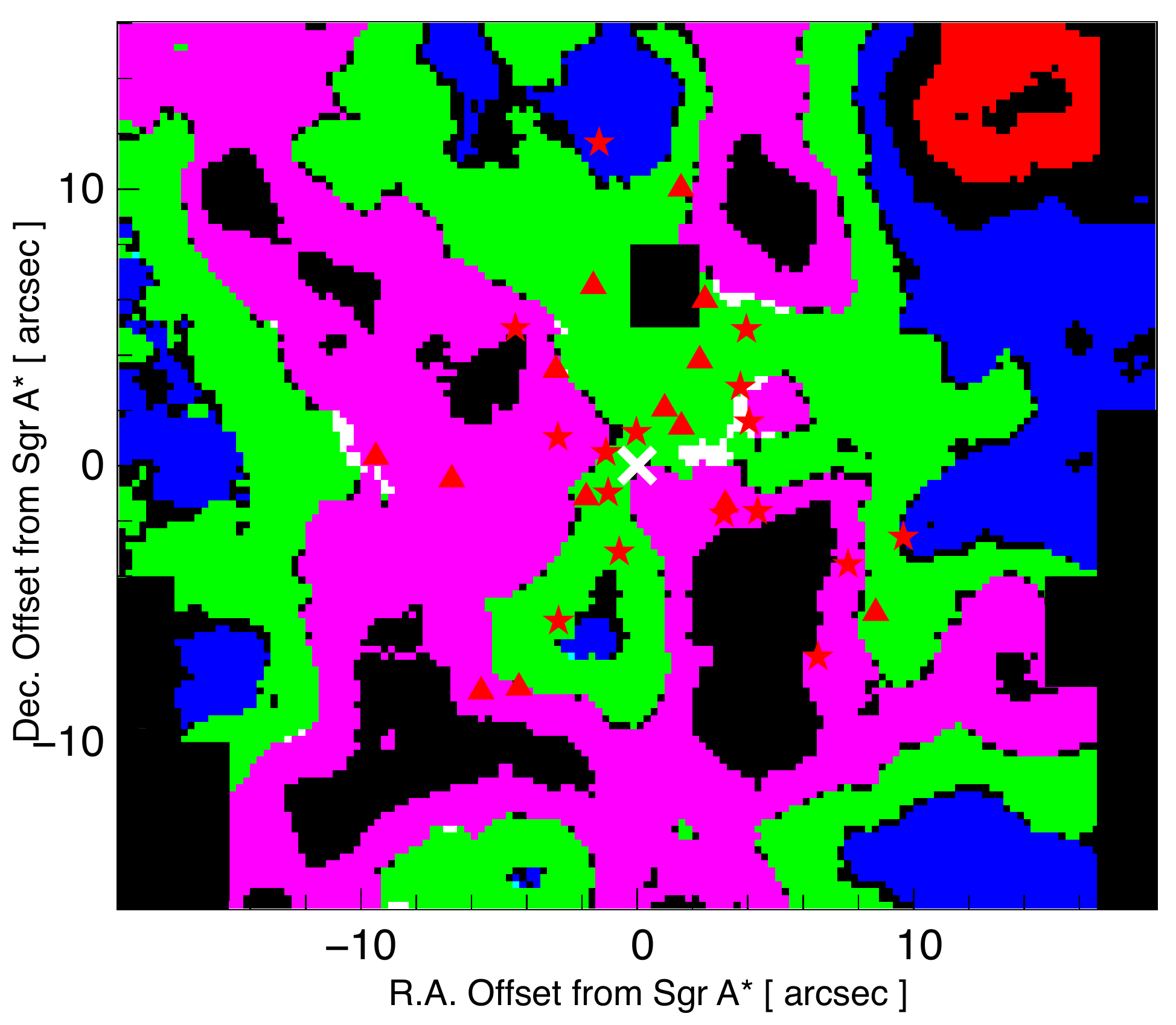}
                \caption{\textit{Top:} As in Fig.~\ref{res:corr}. 
                The groups of correlated points are identified by different colors. \textit{Bottom:} Spatial distribution of the correlated group of points. 
                Each pixel is colored according to the set it belongs to. 
                White pixels indicate where the green and magenta groups overlap and black pixels where there is no coverage by any of the selected groups. 
                Sgr~A* is indicated by the white cross. 
                Red symbols indicate the position of Wolf-Rayet stars: stars for type WN and triangles for type WC (\citealt{Paumard06} catalog). 
                The solid black squared areas indicate regions where we lack observations. GCIRS~7 is covered by the central black square. } 
                \label{res:ind}
                \end{figure}

To investigate these correlations we selected five apparently correlated groups by eye and drew possible correlation curves through them.
Points closer than a chosen distance to the curve are assigned to that group and colored with a tag color. 
The result of this assignment procedure is shown in Fig.~\ref{res:ind} (top).          
This procedure depends on two arbitrary parameters:
(1) the correlation curve, drawn arbitrarily on the distribution of points, chosen to represent the different groups; and (2) 
the maximum distance to the curve, which is chosen in order to cover a maximum of points that seem related to the curve while avoiding any overlap between distinct curves.

This approach does not necessarily address all points and some of them remain black on the plot.
Each colored point of the correlation plot corresponds to a SPIFFI pixel in the dereddened S(1) flux image which can be painted with the same color in the spatial plane. 
The result is a map that shows where the different correlated sets are distributed (Fig.~\ref{res:ind}, bottom). 
Interestingly, each set corresponds to a continuous area on the SPIFFI field and therefore strengthens the idea that the observed correlation is real. 
We tried selecting curves along other random directions to verify our choice of parallel increasing curves. 
This results in more randomly scattered regions in the SPIFFI field of view. 
The combination of almost parallel curves originally chosen is the one that leads to the most contiguous areas on the SPIFFI field of view.

In summary, we observe two effects: extinction correlates with flux and this correlation splits into several distinct areas.
        
\section{Radiative-transfer toy model}
\label{model}
        
If one considers just one of the colored curves obtained in the correlation plot, the problem is to understand how flux and extinction could be correlated. 
This positive correlation means that the larger the extinction, the higher the flux. 
It implies that the emitting material (i.e., the hot molecular gas) is located at the same spatial location as the absorbing material (i.e., dust).
When the total thickness of the medium increases the total amount of photons produced increases, but at the same time the average extinction that affects these photons increases as well. 
The above argument applies for each correlated set of points. 

All correlation curves show the same slope but are shifted toward higher or lower values of the extinction.
An important point is that separated sets correspond to distinct areas on the observed field of view. 
The reason for this separation might reside in the difference in large-scale foreground extinction along the line of sight between each of these areas. 
This extinction acts as a non-emitting ``screen'' which attenuates radiation from the emitting material. 
Each of the painted areas corresponds to a medium which is subjected to a specific foreground extinction.

Starting from these qualitative ideas, we propose a simple phenomenological model to account for the above-described extinction-flux correlation.

To build a phenomenological ``toy'' model we also take into account the fact that we know that the UV radiation is very rapidly absorbed by H$_{2}$ through Lyman and Werner bands. 
This happens at the surface of clumps that are optically thick for UV. 
Therefore, the hot \h~NIR emission arises from a thin shell ($\sim$1/100 of the whole clump length), at the surface of gas clumps \citep{Ciurlo16}, where dust is mixed with the \h~molecules.

\begin{figure}
                \centering
                \includegraphics[width=9cm]{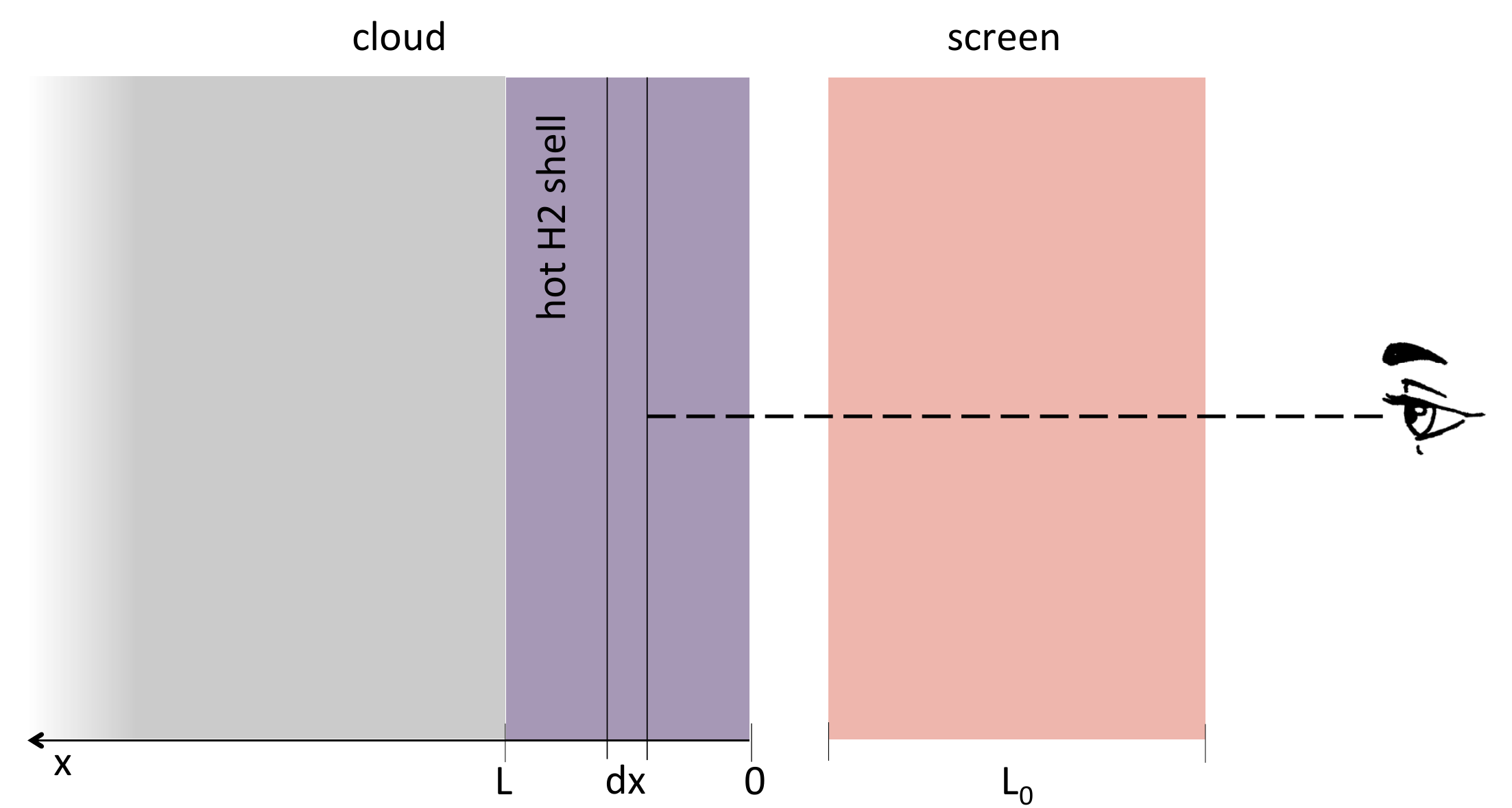}
                \caption{Illustration of the simplified `toy' model representing the emitting medium. 
                The emission comes from a homogeneous shell at the surface of a clump. In this shell,  the emitting material is mixed with the absorbing one. 
                The emission is partially absorbed while traveling across the shell. 
                An additional extinction is caused by a purely absorbing dust screen positioned between the emitting shell and the observer.}
                \label{res:model}
                \end{figure}

As illustrated in Fig.~\ref{res:model}, we assume an emitting shell (at the surface of a clump), composed of homogenous medium.
In the shell, hot gas and dust are mixed and in front of this emitting shell we assume a purely absorbing screen. 
In the medium composing the shell, we assume the excitation temperature to be constant. 
Thus, the optical depth along the geometrical thickness in the medium is given by: 
        \begin{equation}
        \tau_{\lambda}(x)=\alpha_{\lambda} \cdot x, 
        \end{equation}  
where $\alpha_{\lambda}$ is the absorption coefficient per unit of length and $x$ is the path length. 
This extinction is entirely due to dust since self-shielding by \h~itself is totally negligible (see Appendix~\ref{appendix}).
Along the line of sight to the observer, the dust screen absorbs part of the shell emission, adding an optical depth $\tau_{0}$. 
Both optical depths depend on the wavelength $\lambda$ according to the same extinction law.
        
In this case the flux received by the observer is determined by the radiative transfer:
        \begin{equation}
        F_{\lambda} \propto \int_{0}^{L} I_{\lambda}^{0} \cdot e^{-(\tau_{\lambda}(x)+\tau_{0})} ~dx = I_{\lambda}^{0} \frac{ 1-e^{-\alpha_{\lambda} L} }{ \alpha_{\lambda} } \cdot e^{~-\alpha_{\lambda} L_{0}},
        \label{res:Fmod}
        \end{equation}
where $L$ is the thickness of the emitting shell, $L_{0}$ the thickness of the screen, and $I_{\lambda}^{0}$ is the source intensity. 
The shell and the screen have the same extinction coefficient $\alpha_{\lambda}$.
From this equation we can derive line fluxes and extinction.
        
We assume a power law for the NIR extinction curve with an index derived from the mean of various studies as provided by \cite{Fritz11}:
        \begin{equation}
        A_{\lambda} \propto \lambda ^{ -2.07 }
        \label{eq:extlaw}
        .\end{equation}

For a given geometrical thickness, the extinction and extinction coefficients are proportional to each other: 
        \begin{equation}
        \alpha_{\lambda}x = A / (2.5~\log{e}).
        \end{equation}
For 1-0 S(1) and 1-0 Q(3) lines: 
        \begin{equation}
        \alpha_{Q3}/\alpha_{S1}=A_{Q3}/A_{S1}=(\lambda_{Q3}/\lambda_{S1})^{-2.07}=m, 
        \end{equation}
where $m$ is known and depends only on the wavelengths of the two lines.        
We consider all values as relative to the S(1) line and make the notation simplification $\alpha_{S1}=\alpha$ and $\tau_{S1}=\tau$.
It follows that $\alpha_{Q3}=m\cdot\alpha$ and $\tau_{Q3}=m \cdot\alpha L=m \cdot\tau$. 
Therefore, for Q(3) and S(1) lines one has
        \begin{equation}
        F_{Q3} \propto I_{Q3}^{0} \frac{ 1-e^{-m \tau} }{ m \cdot \alpha} \cdot e^{-m \tau_{0}},
        \label{fq3}
        \end{equation}
        \begin{equation}        
        F_{S1} \propto I_{S1}^{0} \frac{ 1-e^{-\tau} }{\alpha} \cdot e^{-\tau_{0}},
        \label{fs1}
        \end{equation} 
and it follows that
        \begin{equation}        
        \frac{F_{Q3}}{F_{S1}} = \frac{I_{Q3}^{0}}{I_{S1}^{0}}\frac{ 1-e^{-m \tau} }{1-e^{-\tau}}\frac{e^{\tau_{0}(1-m)}}{m},
        \label{fr}
        \end{equation} 
where $\tau$ and $\tau_{0}$ are the two parameters of the model. 

Since Q(3) and S(1) arise from the same upper level and are optically thin \citep{Gautier76},  we have \citep{Scoville82}
        \begin{equation}
        I_{Q3}^{0}/I_{S1}^{0}=(\mathcal{A}_{Q3} \cdot \lambda_{S1})/(\mathcal{A}_{S1} \cdot \lambda_{Q3})=1.425,
        \end{equation}
where $\mathcal{A}$ is the Einstein coefficient.
Now, to derive the quantity A$_{Ks}$, we assume no radiative transfer effect and consider that
        \begin{equation}
        \frac{F_{Q3}}{F_{S1}}=\frac{ \mathcal{A}_{Q3} \cdot \lambda_{S1}}{ \mathcal{A}_{S1} \cdot \lambda_{Q3}}\cdot 10^{(A_{\lambda_{S1}}-A_{\lambda_{Q3}})/2.5},
        \end{equation}
meaning that
        \begin{equation}
        A_{S1}-A_{Q3} = 2.5 \cdot log_{10} \left( \frac{1-e^{-m \tau}}{1-e^{-\tau}} \frac{e^{\tau_{0}(1-m)}}{m}\right). 
        \label{E}
        \end{equation}
Equation~\ref{eq:extlaw} and $\lambda_{Ks} = 2.149$~$\mu m$:
        \begin{equation}
        A_{Ks} = \frac{\lambda_{Ks}^{-2.07}}{\lambda_{S1}^{-2.07}-\lambda_{Q3}^{-2.07}} \cdot 2.5 ~ log_{10}\left( \frac{1-e^{-m \tau}}{1-e^{-\tau}} \frac{e^{\tau_{0}(1-m)}}{m} \right).
        \label{aks}
        \end{equation}
         
To summarize, we present an equation for the extinction that can be used to compare the model with the data.
One has to underline once more that A$_{Ks}$ derived in this way is what we would obtain if a foreground screen was attenuating the emission coming from a thin emitting shell of hot, optically thin H$_{2}$, as is explained in Sects.~3.3 and 3.4.4 of \citealt{Ciurlo16}.

                \begin{figure}
                \centering
                \includegraphics[width=9cm]{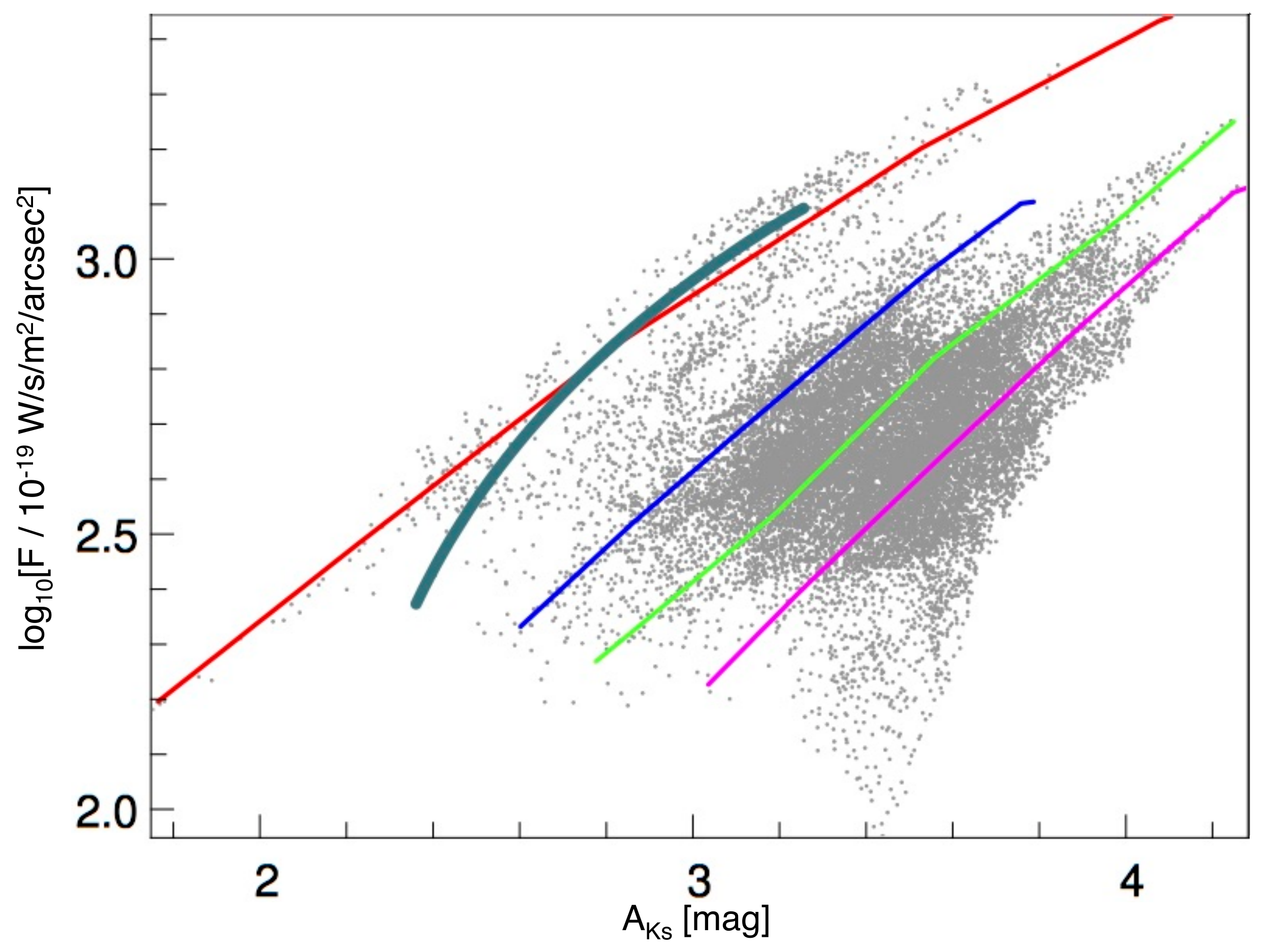}
                \caption{As in Fig.~\ref{res:corr}, with the arbitrarily drawn correlation curves, and the corresponding rigorously determined correlations modeled as in Sect.~\ref{model} (dark green).
                }
                \label{toyModelData}
                \end{figure}

                \begin{figure*}[h]
                \centering
                \includegraphics[width=17cm]{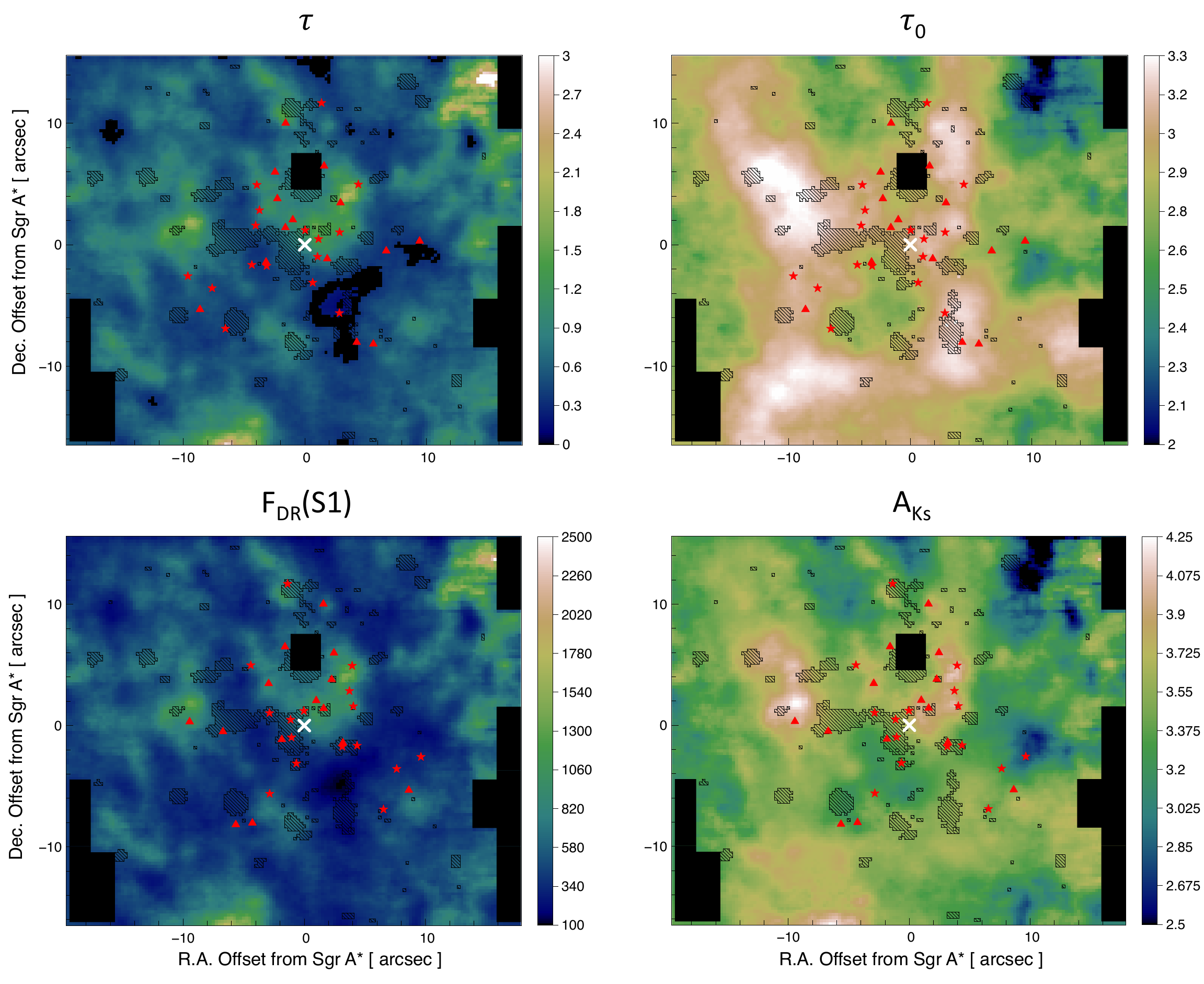}
                \caption{\textit{Top:} 2D representation of $\tau$ (local extinction; \textit{left}) and $\tau_{0}$ (screen extinction; \textit{right}) obtained through the simplified model. 
                These maps closely resemble the dereddened flux (\textit{bottom left}) map and A$_{Ks}$ map (\textit{bottom right}), respectively (bottom images are from \citealt{Ciurlo16}).}
                \label{res:tau}
                \end{figure*}

\section{Discussion}
\label{comparison}

Provided that the model is properly rescaled (see below), Fig.~\ref{toyModelData} shows that the model reproduces the shape of the somewhat arbitrarily chosen correlation curves on Fig.~\ref{res:corr}, especially the slope.
        
Therefore, one can try to invert the problem and compute, for each point (i.e., for each dereddened flux and A$_{Ks}$), the corresponding $\tau$ and $\tau_{0}$. 
This is done by inverting Equation \ref{aks} using the \texttt{newton} IDL routine which solves a system of $n$ nonlinear equations in $n$ dimensions using a globally convergent Newton method\footnote{https://www.harrisgeospatial.com/docs/NEWTON.html} \citep{Press92}. 
To properly compare the model to observations one has to consider a factor which scales the flux obtained through the model in order to match the actual flux measured.
This factor contains the rescaling to the proper flux units, the quantity $I_{Q3}^{0}/I_{S1}^{0}$ and the unknown factor $\alpha$ (i.e., $\alpha_{S1}$ in our notation). 
The value ($0.75~10^{-13}$) has been chosen  because it is the one that allows the Newton procedure to find solutions for the highest number of points. 

Once given the flux factor, the result is a value of $\tau$ and $\tau_{0}$ for each point of the plot (i.e., each pixel on the 2D plane) as shown in Fig.~\ref{res:tau} (top left and top right, respectively). 

The $\tau$ map very closely resembles the dereddened flux distribution (Fig.~\ref{res:tau}, left bottom) while $\tau_{0}$ seems a smoother version of the A$_{Ks}$ map (Fig.~\ref{res:tau}, right bottom). 
This outcome is reasonable since $\tau$ is supposed to represent the extinction within the emitting shell and is thus naturally correlated with flux in this picture. 
The fact that the model reproduces the observed correlation between the dereddened flux and the extinction rather well strengthens this interpretation. 

         \subsection{Foreground extinction} 
        
         Assuming that this interpretation is correct, the offset between the distinct curves in Fig.~\ref{res:ind} (top) corresponds to different foreground screen optical thicknesses (i.e., different $\tau_{0}$) for the various colored areas of Fig.~\ref{res:ind} (bottom).
        
        The differences in screen depth can be interpreted as the geometrical location (along the z-axis) of the emitting medium. 
        For a uniform screen, the further away an object is from the observer, the thicker the screen (i.e., higher values of $\tau_{0}$).
        Indeed, in Fig.~\ref{res:ind} (bottom), the north-western corner of the observed field corresponds to the north-eastern border of the CND.
        According to the known CND inclination \citep{Liszt03}, in the observed field, this is the closest portion of the CND to the observer. 
        Consistently, this area corresponds to the lowest foreground extinction among the correlation curves (Fig.~\ref{res:ind} top, the red curve). 
        On the contrary, the highest $\tau_{0}$ region (magenta and green) encompasses the Northern Arm of the Minispiral, which is believed to be located at greater line-of-sight distances \citep{Paumard04}. 
        In this interpretation, the variation of $\tau_{0}$ is foreground relative to the emitter, but still local to the Galactic center.
        If this interpretation is correct (further corresponds to more absorbed) then it is only very locally that this extinction is effective since these are indeed very small distance variations when compared to the distance of the Galactic center. On a range of 5-pc scale,
$\delta A_K = 1$  translates to $A_V\simeq2$ per parsec. 
        This value, compared to $A_V = 2$ for 1~kpc on average through the Galactic disk, is a factor of 1000 larger. 
        This is a big difference, but is not unrealistic: for instance, a Bok globule is very absorbing but also very compact (e.g., in  \citealt{Alves01}, $A_V = 30$ for a diameter of 0.12~pc).
        Therefore, the above interpretation is plausible. 
        
        However, given the filamentary nature of the ISM, some structure in the foreground extinction is expected and it may account for the observed spatial variations in the screen absorption.
        For instance, some of the foreground extinction could happen in the clumps and filaments of the central molecular zone (CMZ, \citealt{Morris96}).
        The 20 and 50 km/s clouds lie very close to the central parsec in projection \citep{Guesten83}, within 20 pc of SgrA* according to \cite{Molinari11} and \cite{Henshaw16}. 
        Therefore, it appears reasonable that CMZ's material may cause variation of the foreground screen extinction.

        \subsection{Local extinction} 
        \label{2models}
        
                \begin{figure*}[htb]
                \centering
                \includegraphics[width=10cm]{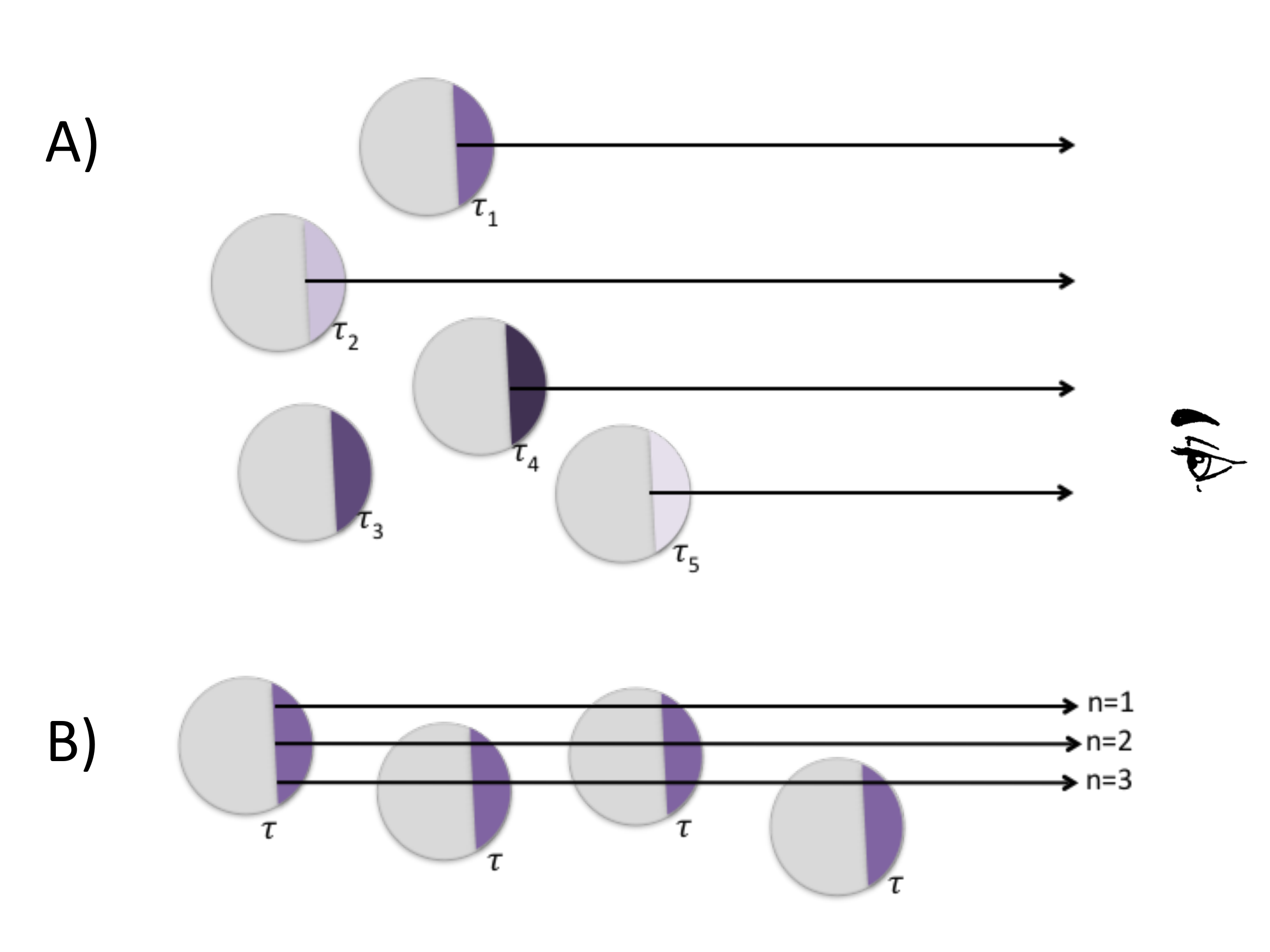}
                \vspace{-0.3cm}
                \caption{
                Illustrations of the two scenarios for the flux-extinction correlation. 
                In both cases, the \h~emission arises from the thin external shell of clumps. 
                In scenario A, these clumps are optically thick to \h~emission and the emitting shells have variable optical depth ($\tau_{1}\neq\tau_{2}\neq\tau_{3}\neq\tau_{4}\neq\tau_{5}$). 
                In this case, only the most external clumps are visible (1, 2, 4, 5, but not 3).
                In scenario B, the clumps are optically thin and are all identical, with optical depth of the emitting shell $\tau=\tau_{C}/100$. 
                In this case, the emission coming from a background clump is still visible, provided that there are not too many clumps on the line of sight.}
                \label{scenarios}
                \end{figure*}
        
\begin{figure}
                \centering
                \includegraphics[width=9cm]{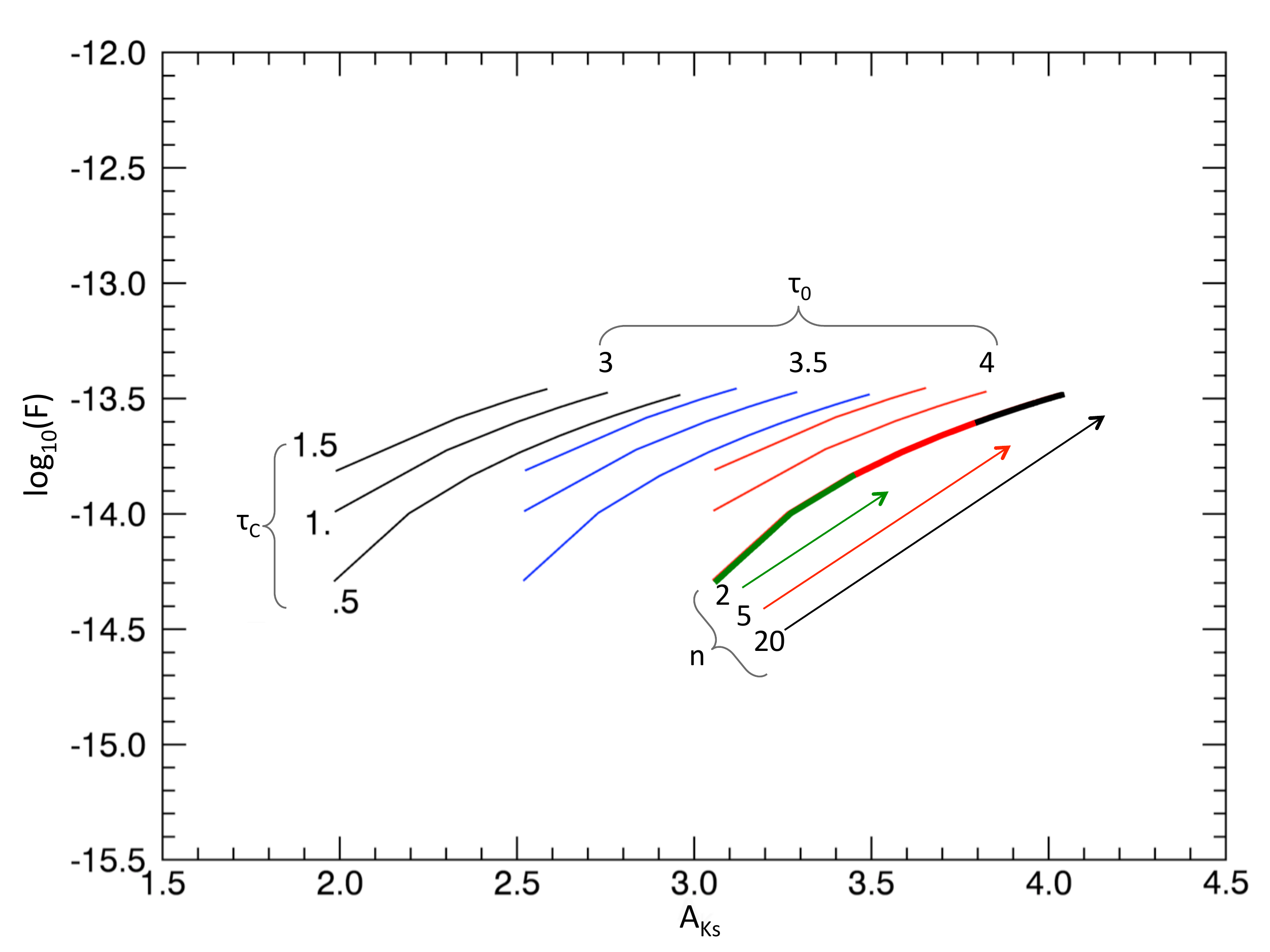}
                \caption{Models from scenario B for different values of the clump absorption coefficient $\tau_{C}$ and different foreground screen depths: $\tau_{0}$=3 (black), 3.5 (blue), 4 (red). 
                The variation of $\tau_{C}$ has the effect of expanding the dynamic range covered by the model. 
                The variation of $\tau_{0}$ shifts the curve to higher extinction values. 
                For $\tau_{0}$=4 the model is traced for 20 (black), 5 (red), and 2 (green) clumps with $\tau_{C}$=0.5 each. 
                Models overlap on the plot and, in fact, all start from the same value. 
                The 20-clump model is the one which covers the whole dynamics. 
                The models reproduce the fact that increasing the number of clumps on the line of sight has the effect of enhancing both the flux and the extinction as observed.}
                \label{Nclouds}
                \end{figure}

        In our model, the flux increases with increasing local extinction because as the physical depth increases the column density of emitting gas increases.
        However, the increase in flux saturates when the clump becomes optically thick to exciting UV radiation.
        To make the optical depth variable, so as to sweep the complete curve of correlation, there are two possible scenarios (see Fig.~\ref{scenarios}):
        
        \begin{itemize}
        
        \item[A)] 
        Clumps optically thick to IR \h~emission: 
        each clump is irradiated differently and therefore each emitting shell has a different optical depth ($\tau$) in the \h~line.
        For increasing optical depth, both emission and extinction are enhanced with the mechanism described in Section \ref{model}. 
        In this scenario only the most external clumps nearest to the observer can be observed (those not hidden behind other clumps). 
        For a photon-dominated region (PDR) in standard conditions, simulations show that the excited \h~is confined to a very thin shell with $r_{shell}/R_{clump}$=1/100 -- 1/1000 (J.~Le~Bourlot, private communication, see also \citealt{Ciurlo16}).
        To reproduce Fig.~\ref{toyModelData}, the optical depth of this emitting shell has to be of the order of 0.5--1. 
        This interpretation implies that the optical depth of the whole clump is very high (A$_{V}$=200--10\,000). 
        This is possible only for very dense clumps.
        \cite{Christopher05} and \cite{Smith14} have shown that such a high density is reached in many clumps composing the CND. 
        \cite{Mills13} finds lower column densities for the CND clumps ($\sim$10$^{5}$ -- 10$^{6.5}$~cm$^{-3}$).
        Therefore, for the CND we considered the clump densities to be in the range 10$^{5}$ -- 10$^{7}$~cm$^{-3}$ and the typical clump size to be of ~0.2~pc \citep{Christopher05}.
        From $n_{H_{2}} /A_V=1.8~10^{21}$ \citep{Predehl95} one derives A$_{V}$=40 -- 4\,000.  
        Therefore, the proposed model is compatible with previous studies.      \\

        \item[B)] 
        Clumps optically thin to IR \h~emission:
        for each clump, the \h~emission is produced in a more or less identical emitting shell of optical depth $\tau$.
        The \h~emission, on its way out of the medium, is attenuated by bulk dust in other clumps but is also enhanced by the emission coming from the emitting shell of these same clumps.
        For an increasing number of clumps on the line of sight, both the emission and the extinction are enhanced (see Fig.~\ref{Nclouds}). 
        The model is the same as described by Equation~\ref{res:Fmod}, but instead of variable absorption coefficient, we considered $n$ separated, identical clumps.
        The flux received by the observer is the sum of all contributions of clouds on the line of sight, each one being partially extinguished by those in front of it, plus the foreground screen:
                \begin{equation} 
                F = [F_0 + F_0 e^{-\tau_C} + F_0 e^{-2 \tau_C} + ... + F_0 e^{- (n-1) \tau_C}]\cdot e^{-\tau_0}.
                \end{equation}
        The first term corresponds to the emitting shell closest to the observer (not extinguished by any cloud).
        The last term corresponds to the most distant shell, which is suffering extinction from all $n-1$ clouds between it and the observer.   
        Since we assume that the \h~emission is confined in an external shell of thickness typically $1/100$ of the size of the cloud $\tau = \tau_C/100$, and $F_0 \propto (1 - e^{- \tau_C/100})$, so that we have:
                \begin{equation} 
                F \propto (1 - e^{- \tau_C/100}) [1 + e^{-2 \tau_C} + ... + e^{- (n-1) \tau_C} ]\cdot e^{-\tau_0}. 
                \end{equation} 
        We consider $\tau_C$ close to $1$ (which is much less than in hypothesis A, where $\tau_C$ was 50--100), so in practice:
                \begin{equation} 
                F \propto  \frac{\tau_C}{100} \cdot [1 + e^{-2 \tau_C} + ... + e^{- (n-1) \tau_C}]\cdot e^{-\tau_0}.  
                \end{equation} 
        
        The final model that best reproduces the observations is the one with from 1 to 20 clumps on the line of sight, each of $\tau_{C}\approx0.5$.
        This model works provided that the filling factor by clumps is not too high, allowing a statistically large variation of the number of clumps from one line of sight to the other. 
        In this case, the photons produced by the different clumps would be attenuated by factors varying in a large range. 
        This scenario would be favored for the central cavity since, there, the filling factor is likely lower than for the CND and the lower medium densities allow for optically thin clumps. \\
                
        \end{itemize}
                        
        \noindent       
        Both scenarios reproduce well the observed slope in the correlation plot.  
        Scenario A seems more applicable to the CND if we consider the very high densities derived by \cite{Smith14, Mills13, Christopher05}. 
        On the other hand, in the central cavity, a high density cannot be attained. 
        There, scenario B seems to be more appropriate even though it still implies fairly high density (20 clumps of A$_{K}\sim1$ are needed to reproduce the observed dynamics, i.e., A$_{K_{cloud}}\sim$20). 
        In practice, the two scenarios probably can coexist in some areas and the final result is likely a superposition of the two effects. 
                
        \subsection{The global extinction effect}
                
\begin{figure*}[h]
                \centering
                \includegraphics[width=18cm]{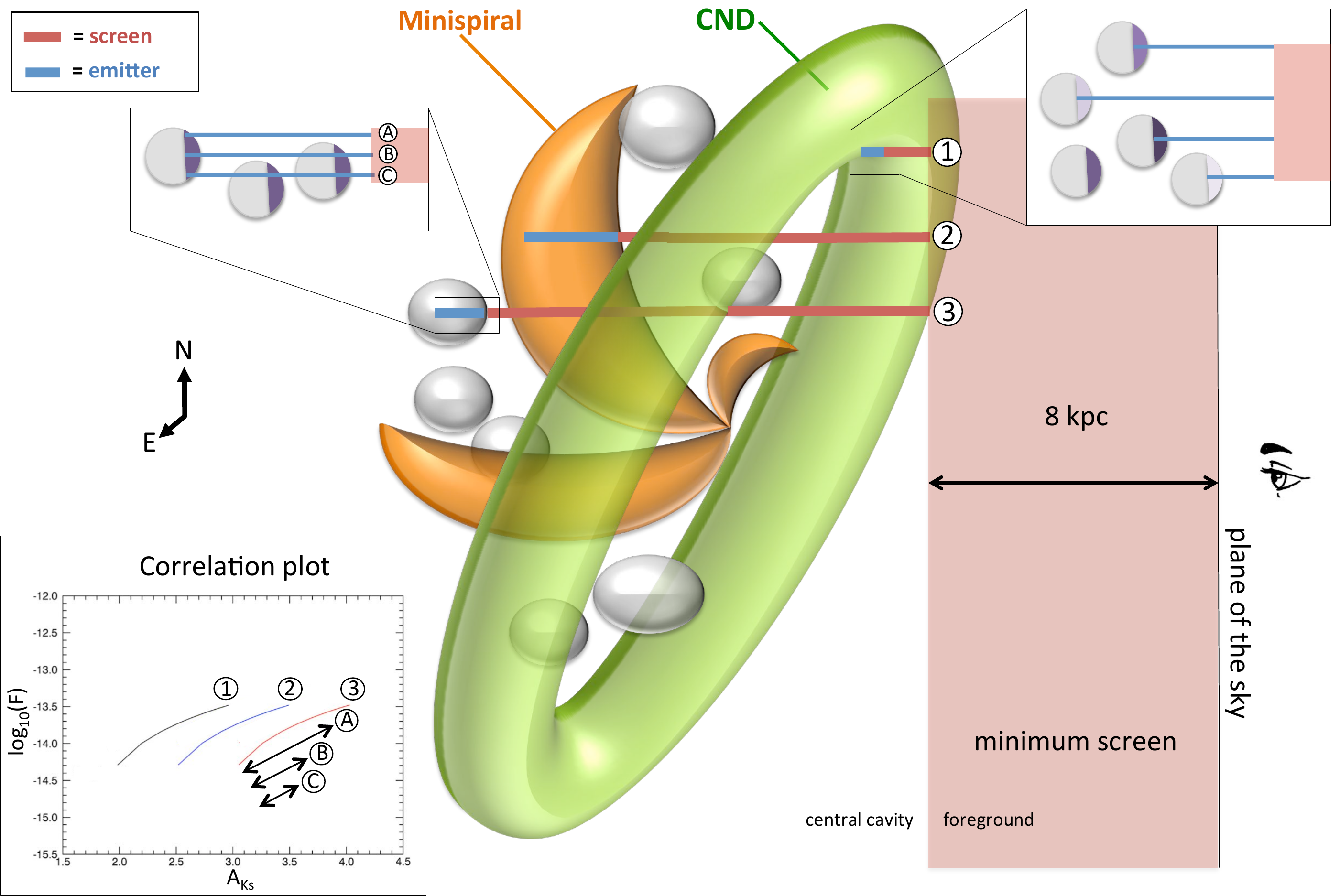}
                \caption{
                Representation of the central parsecs along the  line of sight. 
                The CND is represented in green, the Minispiral in brown.
                The clumps of gas and dust in the central cavity are shown in gray. 
                The foreground, purely absorbant `screen' extinction ($\tau_{0}$) is represented in red, the local ($\tau$) effect of mixed dust and gas (providing for both emission and extinction) is shown in blue. 
                The correlation plot in the bottom-left corner shows the models representing the cases marked on the figure.}
                        \vspace{-10cm}
                \label{3D}
                \end{figure*}

        Figure~\ref{3D} summarizes our interpretation by showing a schematic representation of the central parsec region, seen from the side.
        
        All emitting shells in the region are subjected to the same 8~kpc Galactic disk extinction. 
        Depending on the distance to the observer and the variation of the ISM along the line of sight, each clump is subject to an additional screen-type extinction.
        Case 1 in Fig.~\ref{3D} corresponds to the curve at the lowest extinction on the correlation plot (red curve in Fig.~\ref{res:ind}). 
        Cases 2 and 3 represent the cases where the screen extinction increases.
        
        As previously mentioned, the clumps of the CND are dense and big, meaning that scenario A (Sect.~\ref{2models}) is more representative: each clump is subject to a different irradiation and shows a different optical depth which is represented by the points along the curve of the correlation plot.
        In this representation, the medium is continuous.
                
        In the central cavity, the medium is less dense (i.e., has a smaller filling factor), fragmented in smaller-scale clumps, and scenario B (Sect.~\ref{2models}) is more representative: depending on the number of small clumps on the line of sight the points lay on different parts of the correlation curve (case C in Fig.~\ref{3D} shows the lowest number of clumps).
        A lot of these clumps can overlap on the line of sight. 
        Given the high number of clumps (20 to cover the whole dynamic range on the correlation plot), scenario B actually reproduces the behavior of a continuous medium.

        The two models are therefore not in conflict: they are the extreme representations of two different scale effects. 
        One scenario or the other can apply depending on the scale on which we are observing the ISM. 
        For instance, at the surface of one big clump of scenario A, there could possibly be many much-smaller-scale clumps producing the observed effects, as described in scenario B. 
        We can conclude that, in general, the medium is very clumpy and fragmented.  

\section{Conclusions}
\label{conclusions}

In \cite{Ciurlo16}, for the first time, the Galactic center extinction map was computed directly from the comparison of two molecular lines: H$_{2}$ 1-0 S(1) and 1-0 Q(3). 
Here, we take a step further by interpreting the results in terms of separate effect of the foreground and local extinction. 
The computed extinction appears to be correlated locally with the flux map in several distinct regions. 
Each of these regions is identified by the same slope in the correlation plot. 

We propose two models that successfully reproduce the observed correlation. 
In both models, the correlation is due to the fact that the local emitting H$_{2}$ is mixed everywhere with dust.
With the help of a simple toy model of radiative transfer in a clumpy medium, we showed that in the central parsec around the Galactic center the extinction can be interpreted as a combination of:
        \begin{enumerate}
        \item the effect of a purely absorbing, foreground screen and
        \item the variation of the optical depth of clumps on the line of sight (both because of differences in the optical depth and in the number of emitting shells). 
         \end{enumerate}

Effect $(1)$ causes, for different optical depths of the (purely) absorbing screen, the splitting into multiple curves in the correlation plot. 
Effect $(2)$ causes the variation of the optical depth along the line of sight, and consequently the correlation between the flux and the extinction.
In summary, in the correlation plot, the observed slope traces the fragmentation of the ISM. 
On the other hand, the split into different groups of correlated points translates into variations in the large-scale foreground extinction.
The extinction measured through the Q(3)-S(1) ratio is a combination of both effects. 

These results underline the importance of estimating the extinction for the specific emitter under study since each source is located at a different optical depth for the observer.
Both local and foreground extinction are important to probe: on one side the local structure of the medium, and on the other side the foreground extinction to which our specific emitter is subjected. 

The  central cavity of the CND is far from being empty: diluted and fragmented clumps of excited molecular gas and dust are located there. 
The question of the origin of this material is not yet clear (as discussed in \citealt{Ciurlo16}). 
The presence of dust could allow for local formation of H$_{2}$, provided that the grains are not too hot. 
Even though one can likely imagine that H$_{2}$ is short-lived in this highly ionized environment, a small amount can continuously be produced at the surface of dust grains that are present everywhere. 
The presence of very hot H$_{2}$ is detected in other \ion H{ii} regions as well, such as the Crab Nebula \citep{Scoville82}.
Therefore, it is not so surprising to find some at the Galactic center as well, as was also discussed in \cite{Ciurlo16}.

It must be noted that the observed \h~lines represent only the emission coming from the hot gas.
In the central cavity, there might be more \h~that is not hot enough to be detected in these lines and not dense enough to be detected through other lines (such as HCN).

In summary, the combined study of several NIR \h~lines has allowed us to learn more about the clumpiness of the medium in the central parsec of the Galaxy.
This same kind of study can be applied to other regions where \h~is hot enough to be detected to learn about the structure of the ISM.


\begin{acknowledgements}
We would like to thank the infrared group at the Max Planck Institute for Extraterrestrial Physics (MPE), who built the SPIFFI instrument, carried out the observations, and provided the data.
We also thank the referee for the useful comments, which helped us improve this paper.
\end{acknowledgements}

                
\bibliographystyle{aa} 
\bibliography{aa31763-17}{}

        \onecolumn
\begin{appendix} 
\section{Estimation of the line absorption coefficient in radiative excitation}
\label{appendix}

For a transition from the upper ($i$) to the lower ($j$) level, the absorption coefficient $\alpha_{ji}$ can be calculated as
        \begin{equation}
        \alpha_{ji} = \frac{h~\nu_{ji}}{4\pi}~n_j~B_{ji}~\phi_{ji}, 
        \end{equation} 
where $h$ is the Planck constant, $\nu_{ji}$ is the line frequency (hereafter $\nu$), and $n_j$ the column density (in m$^{-3}$). 
The Einstein coefficient $B_{ji}$ is given by 
        \begin{equation}
        B_{ji}= B_{ij} \frac{g_i}{g_j} = A_{ij} \frac{c^2}{2 h \nu^3}\frac{g_i}{g_j}.
        \end{equation}
$\phi_{ji}$ is the line profile, with $\int \phi_{ji}~d\nu = 1$, and it is given by
        \begin{equation}
        \phi_{ij} = \frac{1}{\Delta \nu} = \frac{c}{\nu \Delta v}= \frac{\lambda_{ij}}{\Delta v},
        \end{equation}
where $\lambda_{ij}$ is the line wavelength and $\Delta v$ the line width.
One can therefore write
        \begin{equation}
        \alpha_{ij} = n_j \frac{\lambda_{ij}^3}{8\pi} A_{ij} \frac{g_i}{g_j} \frac{1}{\Delta v}.
        \end{equation}
The column density can be expressed as $n_{j} = n_{H_{2}}\cdot f_{j}$, where $f_{j}$ is the relative population of the j-level and is given by
        \begin{equation}
        f_j=\frac{e^{-E_j / kT}}{\sum e^{-E_k / kT}}.
        \end{equation}

For 1-0  S(1) ($J=3\rightarrow 1$ , $v=1\rightarrow 0$) at $\lambda  = 2.12~\mu$m: $A_{ij} = 3.48~10^{-7}$ s$^{-1}$, $g_i = 21$, $g_j = 9$ and $E_j = 6.952~10^{3}$~K.
For a PDR such as the central parsec $\Delta v \sim100$~\kms \citep{Ciurlo16}, $n_{H_{2}}=10^{5}$~cm$^{-3}$ \citep{Smith14} and T=1\,200~K as estimated for Orion in \cite{Beckwith83}. 
Therefore one finds $\alpha_{ij}\sim2.7~10^{-22}$~m$^{-1}$. 

Since $\tau_{ij}= \alpha_{ij}\cdot L$ and a typical clump has a diameter of $L = 0.2$~pc \citep{Christopher05}, one finds $\tau_{ij}=1.6~10^{-6}$. 
This value has to be compared to $A_V = n_{H_{2}} \cdot L/1.8~10^{21}$ \citep{Predehl95}. 
For $n_{H_{2}}=10^{5}$~cm$^{-3}$ it gives $\tau_{ij} / A_V = 5~10^{-8}$ which translates to $\tau_{ij} / A_K   \sim 5~10^{-7}$. 
Therefore, the \h~line is optically thin and there is no self-shielding. The extinction is thus completely due to the dust component of clumps.

\end{appendix} 
\end{document}